\documentclass[11pt]{article}
\usepackage[english]{babel}
\usepackage{amsmath,amsfonts,amssymb,amstext,amsmath,amsthm,amscd}
\usepackage{mathtools}
\usepackage[a4paper,top=1.5cm,bottom=2cm,left=2cm,right=2cm]{geometry}
\usepackage{breakcites}

\usepackage{booktabs}
\usepackage{mathtools}
\usepackage{mathrsfs}
\usepackage{dsfont}

\usepackage{graphicx}
\usepackage{subfigure}
\usepackage{wrapfig}
\usepackage{indentfirst}
\usepackage{latexsym}
\usepackage{sidecap}
\usepackage{booktabs}
\usepackage{hyperref}

\usepackage{lipsum}
\usepackage{cleveref}
\usepackage[dvipsnames]{xcolor}

\newcommand\myshade{85}
\colorlet{mylinkcolor}{violet}
\colorlet{mycitecolor}{YellowOrange}
\colorlet{myurlcolor}{RoyalBlue}

\hypersetup{
    colorlinks=true,
    linkcolor=myurlcolor,
    citecolor=mycitecolor!\myshade!black,
    filecolor=magenta,      
    urlcolor=magenta,
}
\usepackage{setspace}

\usepackage{mathrsfs}
\usepackage{textcomp}
\usepackage{braket}
\usepackage{colortbl}
\usepackage{bbm}
\usepackage{comment}
\usepackage[colorinlistoftodos]{todonotes}\usepackage{mathrsfs}
\usepackage{textcomp}
\usepackage{braket}
\usepackage{colortbl}
\usepackage{bbm}
\usepackage{comment}
\usepackage[colorinlistoftodos]{todonotes}

\usepackage{enumitem}

\usepackage{breakcites}

\newcommand*{\addheight}[2][.5ex]{%
  \raisebox{0pt}[\dimexpr\height+(#1)\relax]{#2}}

\makeatletter
\newcommand\xdleftrightarrow[2][]{\ext@arrow 0099{\longleftrightarrowfill@}{#1}{#2}}
\def\longleftrightarrowfill@{\arrowfill@\leftarrow\relbar\rightarrow}
\makeatother

\theoremstyle{plain}
\newtheorem{Theorem}{Theorem}[section]
\theoremstyle{definition}

\theoremstyle{remark}
\newtheorem{Remark}[Theorem]{Remark}
\theoremstyle{remark}

\theoremstyle{plain}

\theoremstyle{plain}

\theoremstyle{plain}

\theoremstyle{remark}

\theoremstyle{remark}

\theoremstyle{remark}

\theoremstyle{remark}

\newcommand{\XX}{\mathcal{X}}

\newcommand{\red}[1]{\textcolor{red}{#1}}

\DeclareMathAlphabet{\pazocal}{OMS}{zplm}{m}{n}

\newcommand{\YY}{\mathcal{Y}}
\newcommand{\EE}{\mathcal{E}}

\usepackage{tikz}
\usetikzlibrary{arrows}

\usepackage{tikz}
\usetikzlibrary{arrows}

\makeatletter
\pgfdeclareshape{datastore}{
  \inheritsavedanchors[from=rectangle]
  \inheritanchorborder[from=rectangle]
  \inheritanchor[from=rectangle]{center}
  \inheritanchor[from=rectangle]{base}
  \inheritanchor[from=rectangle]{north}
  \inheritanchor[from=rectangle]{north east}
  \inheritanchor[from=rectangle]{east}
  \inheritanchor[from=rectangle]{south east}
  \inheritanchor[from=rectangle]{south}
  \inheritanchor[from=rectangle]{south west}
  \inheritanchor[from=rectangle]{west}
  \inheritanchor[from=rectangle]{north west}
  \backgroundpath{
    \southwest \pgf@xa=\pgf@x \pgf@ya=\pgf@y
    \northeast \pgf@xb=\pgf@x \pgf@yb=\pgf@y
    \pgfpathmoveto{\pgfpoint{\pgf@xa}{\pgf@ya}}
    \pgfpathlineto{\pgfpoint{\pgf@xb}{\pgf@ya}}
    \pgfpathmoveto{\pgfpoint{\pgf@xa}{\pgf@yb}}
    \pgfpathlineto{\pgfpoint{\pgf@xb}{\pgf@yb}}
 }
}
\makeatother

\begin{document}

\title{The Generalized Stochastic Microdosimetric Model: the main formulation}
\author{F. Cordoni\,$^{1,2} $, M. Missiaggia\,$^{2,3}$, A. Attili\,$^4 $, S. M. Welford$^{5} $,  E. Scifoni\,$^2$, C. La Tessa\,$^{2,3}$}
\date{}
\maketitle

\renewcommand{\thefootnote}{\fnsymbol{footnote}}
\footnotetext{{\scriptsize $^1 $University of Verona, Department of Computer Science, Verona, Italy.}}
\footnotetext{{\scriptsize $^2 $TIFPA-INFN, Trento, Italy.}}
\footnotetext{{\scriptsize $^3 $University of Trento, Department of Physics, Trento, Italy.}}
\footnotetext{{\scriptsize $^4 $INFN, Roma 3, Italy.}}
\footnotetext{{\scriptsize $^5 $University of Miami, Miller School of Medicine, Department of Radiation Oncology, Miami, FL, USA.}}

\begin{abstract}
The present work introduces a rigorous stochastic model, named Generalized Stochastic Microdosimetric Model (GSM$^2$), to describe biological damage induced by ionizing radiation.
Starting from microdosimetric spectra of energy deposition in tissue, we derive a master equation describing the time evolution of the probability density function of lethal and potentially lethal DNA damage induced by a given radiation in a cell nucleus. The resulting probability distribution is not required to satisfy any \textit{a priori} assumption.
After the initial assumption of an instantaneous irradiation, furtherrmore, we generalized the master equation to consider damage induced by a continuous dose delivery.
In addition, spatial features and damage movement inside the nucleus have been taken into account. 
In doing so, we provide a general mathematical setting to fully describe the spatiotemporal damage formation and evolution in a cell nucleus.
Finally, we  provide numerical solutions of the master equation exploiting Monte Carlo simulations to validate the accuracy of GSM$^2$.  Development of GSM$^2$ can lead to improved modeling of radiation damage to both tumor and normal tissues, and thereby impact treatment regimens for better tumor control and reduced normal tissue toxicities.
\end{abstract}

\textbf{Keywords or phrases: } Radiation Biophysical Modeling, Microdosimetry, Charged Particle Beam Radiation, Radiation Damage.

\maketitle

\section{Introduction}


Currently, around $50 \%$ of all patients with localized malignancies undergo treatment including ionizing radiation, mostly in combination with tumor resection and/or chemotherapy \cite{Durante2,ptcog}. Conventional therapy with high energy photons is by far the most common approach, but the use of accelerated particles has grown exponentially, especially in the past decade. The well-defined, energy dependent, range with sharp distal fall-off and the limited lateral beam spread, typical of ions when penetrating a medium, translate into a dose profile delivered with millimeter precision. In addition, charged particles,especially for larger charge, present enhanced biological effectiveness compared to photons, resulting in reduced cellular repair \cite{Durante,Sch,Tin}.  Thus the field of radiation oncology is evolving towards broader application of radiotherapy with ions, while still several key physical and biological questions remain to be fully unraveled.
In particular, the need to account for a biologically effective dose, beyond a purely physical energy deposition imposes an advanced  characterization of a beam, \cite{Kramer}.
The calculation of the effective dose distribution delivered to the patient during a treatment, indeed, requires detailed knowledge of the radiation field composition at the tumor site and surrounding tissue. The beam quality across its propagation in the medium is, in fact, modified by nuclear and electromagnetic interactions of the primary ions with the patient's body nuclei, atoms and molecules, creating a mixed radiation field composed of primary as well as secondary nuclear fragments of different charge and kinetic energy, \cite{Durante,Kra}.
Such a complex field, contribute with different components delivering  a different nanoscopic damage to the biological target molecules,  mainly mediated by their secondary electrons distribution, \cite{Sci,Plan}, and on their  turn by the radicals generated by the latter electrons, \cite{Plan2,Plan3}, despite alternative processess, \cite{Sur2,Tou}. Such a nanoscopic pattern of energy deposition, returns in a different \textit{complexity} of molecular damage, \cite{Sur}, which correlates with a different repairability and thus a different biological response.
An accurate approach for characterizing the complex radiation field produced by an ion beam is microdosimetry \cite{Ros}. There are two main points of strength in using this methodology: i) the energy deposited by radiation is measured in an area with dimensions comparable to a cell nucleus; and ii) stochastic fluctuations of energy deposition, e.g. from cell to cell, are taken into account. Microdosimetry is considered a link between the physical characteristics and the biological effectiveness of a radiation with the advantage of an experimentally measurable physical quantity, and has been used in radiobiological models to describe radiation quality. One of the most relevant examples is the \textit{Microdosimetric Kinetic Model} (MKM), which was formulated in its original version in \cite{Haw,Haw3} as an elaboration of the \textit{Theory of Dual Radiation Action} (TDRA)\cite{Kel,Zai} and of the \textit{Repair--Misrepair Model} (RMR) \cite{Cur,Tob}. The MKM exploits microdosimetric spectra to calculate the energy deposited by radiation and predicts cell survival modeling the DNA-damage repair kinetics. Today, it is one of two radiobiological models employed clinically in particle therapy, together with the local effect model (LEM), \cite{Els,Pfu}.

Although based on microdosimetry, the MKM is a purely deterministic model as only the average number of lethal lesions induced by radiation to the DNA is considered. The model aims to provide a mathematical formulation of the kinetic evolution of double--strand breaks (DSB) in the DNA in order to calculate the cell survival fraction. Mathematically, the temporal evolution of a DSB is described by a system of two ordinary differential equations representing the average number of lethal and potentially lethal damages as a function of time. This description is accurate only as long as the lethal and potentially lethal damage distributions are Poissonian, and results in a cell survival curve that follows a linear--quadratic behaviour \cite{Haw}. However, it has been widely shown in literature that the DNA damage distribution deviates significantly from a Poisson function under several irradiation conditions, such as high--dose or high--LET \cite{Haw2}. For this reason, several recent studies focused on implementing corrections to the original MKM formulation to account for non--Poissonian behaviours of the DNA damage distribution, \cite{Haw2,Haw3,Kas,Ina,Ina2,Sat}. An extensive collection of the original MKM formulation and its subsequent generalizations can be found in \cite{Bel}. Nonetheless, all MKM versions are based on the original deterministic formulation described by Hawkins \cite{Haw}.

The main goal of the present work is to develop a fully probabilistic model of DNA damage formation and its kinetic evolution based on microdosimetry. The new model, called (\textit{Generalized Stochastic Microdosimetric Model} (GSM$^2$), will provide a rigorous and general mathematical description of DNA damage time--evolution without using any a priori assumption on the lesion distribution (e.g. a Poisson). The model accuracy will be tested for different irradiation conditions (beam quality, dose and dose rate) and compared with MKM predictions, to prove both GSM$^2$ validity and advances compared to the current standard.

The classical approach for mathematically modeling a complex physical system, such as the one resulting from the interaction between cells and ionizing radiation that leads to the formation of DNA lesions, is achieved with deterministic models. In these approaches, given an initial condition the system time--evolution can be completely characterized at each state. Recent studies \cite{Smi} have shown that this approach fails mainly for three reasons: i) a precise and accurate estimation of the parameters if often non feasible; ii) it is unrealistic to account for all possible interactions as the system complexity increases; and iii) certain systems can be over-sensitive to some input parameters, typically the initial values. All above reasons have led to inclusion of stochasticity in the models via suitable random variables.

To model complex physical processes, such as lesions formation following a radiation exposure, the standard method is to consider the \textit{macroscopic} system, so that the main focus is on the system as a whole; this approach typically yields that the main equations governing the physical or biological processes are deterministic representing average values. In a \textit{microscopic} (or often nanoscopic) approach, instead, each element of the system is usually modelled using Brownian dynamics \cite{VK,Sol,Ianik}. However, the complexity of lesion formation and time-evolution makes a full Brownian dynamics-representation not feasible. 

To obtain a more general and accurate description of DNA lesion formation and evolution than the one provided by a \textit{macroscopic} approach, and yet to maintain suitable mathematical tractability of the main equations, which is often missing in a \textit{microscopic} approach, a hybrid methodology, known as \textit{mesoscopic}, can be considered. This approach takes into account the stochastic nature of a system while remaining manageable from both the analytical and numerical points of view. The \textit{mesoscopic} method is based on the assumption that the process driving the system evolution is a Markov jump process \cite{Gar}. The equations of motion are described via the so--called \textit{master equation} that contains the probability density function of the whole system \cite{Gar,VK,Web}.

In GSM$^2$, we will introduce an equation, referred to as \textit{Microdosimetric Master Equation} (MME), that governs the time evolution of the joint probability density function for lethal and sub--lethal damages inside the cell nucleus and is based on the parameters $a$, $b$ and $r$. The main innovation with respect to the existing approaches is that in the proposed MME they are taken into account variations in both lesions formation and evolution caused by the randomness of these processes. In particular, we will use microdosimetry spectra for describing radiation quality and considering the stochastic nature of energy deposition.

To provide a rigorous mathematical formulation of the DNA damage kinetics, we will consider lethal and sublethal lesions inside a single cell nucleus. Potentially lethal lesions can either be repaired or not, in which case they become lethal lesions. A cell in which at least one lethal lesion has been formed is considered inactivated. A potentially lethal damage induced by radiation can undergo three main processes: (i) it can spontaneously repair at a rate $r$; (ii) it can spontaneously become a lethal damage  at a rate $a$; or (iii) it can combine with another potentially lethal lesion to form a lethal lesion at rate $b$.

Starting from some probabilistic assumptions on the lesions formation, we will derive a master equation that describes the time evolution for the joint probability density function of DNA lesions, both lethal and potentially lethal. The density function solution will be shown to have first moment in agreement with the standard MKM driving equations. The main goal of this study is to overcome the Poissonian assumption on lethal lesions.

In the present work, we will further generalize the MME in two main directions. In particular, besides the damage kinetic mechanisms (i), (ii) and (iii) introduced above, we will additionally consider that: (iv) either a lethal or sub--lethal damage can be formed randomly due to the effect of the ionizing radiation at a rate $\dot{d}$; and (v) lethal lesions can move inside the cell nucleus. Case (iv) represents DNA damage formation resulting from a continuous irradiation field. In fact, together with standard lesion interactions, we will also take into account random jumps in the number of lethal and sub--lethal lesions caused by the stochastic nature of energy deposition.

Case (v), instead, accounts for the the fact that we allow lesions to move between adjacent domains. Because GSM$^2$ model consider pair--wise interactions of potentially lethal lesions, the domain size plays a crucial role. In fact, a domain too big implies that lesions created far away from each other can interact to form a lethal lesion. On other hand, a domain too small results in a lower number of lesions per domain so that the probability of double events can be underestimated. In the limit case, the domain size approaches zero, most domains contain a single lesions and interactions cannot occur \cite{Isa,Hel}. To minimize the model dependence on the domain size, we will allow interactions between lesions both belonging to the same domain and to different domains \cite{Smi}.

In summary, we will introduce a general master equation that models the joint probability distribution of DNA lethal and potentially lethal lesion inside a cell nucleus. The derived master equation will consider, besides potentially lethal lesion repair and death due to either spontaneous dead or pair--wise interaction, also the stochastic effect of energy deposition due to ionizing radiation and lesions movements between adjacent domains, providing a global description of the cell nucleus as a whole. To validate GSM$^2$, we will consider microdosimetric energy spectra obtained from Geant4 simulations \cite{G4}. We will show how different assumptions related to the probability distribution of damages number, as well as model parameters, show significant deviation from the Poisson distribution assumed by all existing models, including the MKM. We will further compute the survival probability and compare it to the classical \textit{linear--quadratic} (LQ) model \cite{Bod,McM}.

The innovations presented in this work are several. We will develop a fully probabilistic description of the DNA damage kinetic. In particular, the joint probability distribution of the number of sub-lethal and lethal lesions will be modelled. We will further generalize the model including inter--domain movements and continuous damage formation due to protracted dose. The resulting \textit{master equation} solution will provide the real probability distribution without any \textit{a priori} assumption on the density function, allowing to compute several biological endpoints. The proposed approach will be able to fully describe the stochastic nature of energy deposition both in time and space, improving the existing models were the energy deposition is averaged over both the whole cell nucleus and cell population. In doing so, we will be able to reproduce several behaviours referred to in literature as \textit{non--Poissonian effects}, that cannot be predicted by the MKM and its variants and are typically included in the models with ad hoc corrections \cite{Kas,Sat,Haw2,Haw3}.

Because of GSM$^2$ flexibility and generality, analytical solutions both on the probability density function and on the resulting survival curve are not of easy derivation. Therefore, the present study is intended as a first step of a systematic investigation of the stochastic nature of energy deposition and how it influences lesion formation. In particular, a further investigation will focus on long--time behaviour of the \textit{master equation} and the resulting survival curve. Furthermore, the principles used in the current approach will be used to develop a fully stochastic model of inter-cellular damage formation optimized to improved radiation field characterization via a novel hybrid detector for microdosimetry, \cite{Mis}.

With GSM$^2$ and its future developments, we try to shed a new light on non--Poissonian effects, to obtain a deeper mechanistic understanding which will allow us to model them more accurately.

The present paper is structured as follows: Section \ref{SEC:MKM} recalls basic assumptions and formulation of the MKM model and its variants, \cite{Sat,Ina,Ina2,Haw3,Haw2}. Then Section \ref{SEC:ME} introduces the main \textit{master equation} describing the probability distributions of lethal lesions. Subsection \ref{SEC:InitD} shows in details how microdosimetric spectra can be used to extract the energy deposition. Thus, Subsections \ref{SEC:Split}-\ref{SEC:Vox} introduced the above mentioned generalization of the \textit{master equation} to consider split dose and domain interconnection. Connection of the current model to the standard MKM are presented in Subsection \ref{SEC:Conn}. Further, long--time behaviour and survival probability resulting from the GSM$^2$ are presented in Section \ref{SEC:Surv}. At last Section \ref{SEC:Num} presents some numerical examples aiming at highlighting specific aspects resulting from the governing \textit{master equation}.

\section{Fundamentals on the Microdosimetric Kinetic Model and related non--Poissonian generalizations}\label{SEC:MKM}

The Microdosimetric Kinetic Model (MKM) is based on the following assumptions:
\begin{enumerate}
\item the cell nucleus can be divided into $N_d$ independent domains;
\item radiation can create two different kinds of DNA damage, referred to as type $I$ and $II$;
\item type $II$ lesions cannot be repaired, and for this reason will be also called lethal lesions. On the contrary, type $I$ lesions, also called sublethal, can be either repaired or evolve into a lethal lesions either by spontaneous dead or by interaction with another sublethal lesion;
\item the number of type $I$ and $II$ lesions in a single domain $d$ is proportional to the specific energy $z$ delivered by radiation to the site;
\item cell death occurs if at least one domain suffers at least one lethal lesion.
\end{enumerate}

In the described setting, lethal lesions represent clustered double-strand breaks that cannot be repaired whereas sublethal lesions are double-strand breaks that can be repaired. 

Denoting by $\bar{x}_{g,z_d}$ and $\bar{y}_{g,z_d}$ the average number of type $II$ (sub--lethal) and type $I$ (lethal) lesions, respectively, induced in the domain $d$ that received a specific energy $z_d$, the following set of coupled ODE is satisfied
\begin{equation}\label{EQN:LQM}
\begin{cases}
\frac{d}{dt} \bar{y}_{d,z_d}(t) =  a \bar{x}_{d,z_d} + b \bar{x}_{d,z_d}^2\,,\\
\frac{d}{dt} \bar{x}_{d,z_d}(t) = - (a+r) \bar{x}_{d,z_d} - 2 b \bar{x}_{d,z_d}^2\,.\\
\end{cases}
\end{equation}

Assuming further that $(a+r) \bar{x}_{d,z_d} >> 2b \bar{x}_{d,z_d}^2$, equation \eqref{EQN:LQM} can be simplified as
\begin{equation}\label{EQN:LQM2}
\begin{cases}
\frac{d}{dt} \bar{y}_{d,z_d}(t) =  a \bar{x}_{d,z_d} + b \bar{x}_{d,z_d}^2\,,\\
\frac{d}{dt} \bar{x}_{d,z_d}(t) = - (a+r) \bar{x}_{d,z_d}\,.\\
\end{cases}
\end{equation}
For ease of notation, we will omit the subscript $(d,z)$ and indicate $\bar{x}_{d,z}:=\bar{x}$ and $\bar{y}_{d,z}:=\bar{y}$.

One of the main goals of the MKM model, is to predict the survival probability of cell nuclei when exposed to ionizing radiation, whose quality is described with a microdosimetry approach. In order to achieve this result, an additional assumption to those listed above must be made:

\begin{enumerate}[resume]
\item lethal lesions follows a Poissonian distribution.
\end{enumerate}

Under the latter assumption, the probability  $S_{d,z_d}$ that a domain $d$ survives as $t \to \infty$ when receiving the specific energy $z_d$, can be computed as the probability that the random outcome of a Poisson random variable is null. Therefore, $S_{d,z_d}$ is given by
\begin{equation}\label{EQN:SurvMKMD}
S_{d,z_d} = e^{-\lim_{t \to \infty} \bar{y}_{d,z_d}(t)}\,.
\end{equation}
The explicit computation \cite{Haw,Man} shows that the number of lethal lesions as $t \to \infty$ can be expressed as
\begin{equation}\label{EQN:Lim1}
\lim_{t \to \infty} \bar{y}_{d,z_d}(t) = \left (\lambda + \frac{a \kappa}{a+r}\right )z_d + \frac{b \kappa^2}{2(a+r)} z_d^2\,,
\end{equation}
Combining equation \eqref{EQN:Lim1} and \eqref{EQN:SurvMKMD} we obtain
\[
S_{d,z_d} = e^{-A z_d - B z_d^2}\,,
\]
with $A$ and $B$ some suitable constants independent of $d$ and $z_d$.

The survival probability \eqref{EQN:SurvMKMD} can be extended to the whole cell nucleus ($S_n$), by averaging it on all domains as
\begin{equation}\label{EQN:SurvMKMC}
\begin{split}
S_{n,z_n} &:= \exp \left (-\sum_{d=1}^{N_d} \langle \lim_{t \to \infty} \bar{y}_{d,z_d}(t) \rangle \right ) \,.
\end{split}
\end{equation}

At last, by averaging $S_{n,z_n}$ over the entire cell population, the overall cell survival can be calculated as:

\begin{equation}\label{EQN:SurvMKMPFin}
S = exp\left (\alpha D + \beta D^2\right )\,.
\end{equation}
where $D$ is the macroscopic dose delivered to the entire cell population.

Details on how the survival function $S$ was derived can be found in \cite{Haw,Haw2,Haw3,Kas}.

Several generalizations \cite{HawIna2,HawIna,Ina,Ina2,Haw2,Sat,Kas} have been proposed to take into account effects due to a deviation of the lethal lesion behavior from a Poissonian distribution. All models try to correct the survival probability \eqref{EQN:SurvMKMPFin} introducing some correction term based on the overkilling effects. By overkilling effect it is intended when a single particle deposits much more energy than is required to kill a cell, \cite{Cha}, so that it kills less cells per absorbed dose. The typical survival correction is of the form, \cite{Haw2},
\[
 S = \exp\left ( -(\alpha_0 +  f(\bar{z}_d,\bar{z}_n)\beta)D - \beta D^2\right )\,,
\]
where $f(\bar{z}_d,\bar{z}_n)$ is a suitable correction term that depends on both energy deposition on the single domain ($\bar{z}_d$ ) and on the cell nuclues ($\bar{z}_n$). An alternative form is given by \cite{Kas}
\[
 S = \exp\left ( -(\alpha_0 + \bar{z}_d^* \beta) D - \beta D^2\right )\,,
\]
where $\bar{z}_d^*$ is a term that accounts for the overkilling effects.

We refer to \cite{Bel} for a comprehensive review of the biophysical models of DNA damage based on microdosimetric quantities.

It is worth highlighting that all corrections so far proposed for non--Poissonian effects rely on ad hoc terms derived from empirical considerations. The final goal of this study, instead, is to obtain analogous corrections based on physical considerations stemming from the stochastic nature of energy deposition, \cite{Loa}.

\section{The Generalized Stochastic Microdosimetric Model GSM$^2$}\label{SEC:ME}

As a part of this study, we investigated how the models described in Section \ref{SEC:MKM} could be developed to rely on the whole probability distribution rather than simmply on its mean value. In fact, all proposed generalizations of the MKM always consider deterministic driving equations for predicting the number of lethal and sub-lethal lesions. Non-Poissonian effects are often proposed as corrections terms added to the survival fraction predicted by the MKM with no formal mathematical derivation and mainly based on empirical evaluations.

The MKM formulation is based on the probability distribution of inducing a damage when a specific energy $z$ is deposited. Once the survival for a given $z$ is computed, the specific energy is averaged over the whole cell population to yield the overall expected survival probability. To the best of our knowledge, there is no systematic investigation that aims at capturing the true stochasticity of both energy deposition and lesion formation.

The main goal of the present work is thus to generalize microdosimetric based models in order to describe the full probability distribution of lethal and sub-lethal lesions. We will take advantage of assumptions $1-5$ described in Section \ref{SEC:MKM}. Regarding assumption $4$, the MKM assumes that the lethal lesions initial distribution, given an energy deposition $z$, follows a Poisson law. We will generalize this assumption assuming a general initial distribution, allowing to fully describe the stochastcic nature of energy deposition. This point will be treated in detail in Section \ref{SEC:InitD}.

An additional remark on the importance of the initial distribution is necessary to fully understand the implication of the generalization we will carry out in this study. The stochasticity of energy deposition in a microscopic volume is the basic foundation of microdosimetry, and assuming every probability distribution to be Poissonian is a restrictive assumption that limits the model application. 

In order to capture the real stochastic nature of energy deposition and related DNA damage formation we will provide a probabilistic reformulation of equation \eqref{EQN:LQM}. We denote by $\left (Y(t),X(t)\right )$ the system state at time $t$, where $X$ and $Y$ are two $\mathbb{N}_0-$valued random variables representing the number of lethal and sub--lethal lesions, respectively. We will consider a standard complete filtered probability space $\left (\Omega,\mathcal{F},\left (\mathcal{F}_t\right )_{t \geq 0},\mathbb{P}\right )$ that satisfies the usual assumptions of right--continuity and saturation by $\mathbb{P}-$null sets.

Let us consider two different sets $\XX$ and $\YY$ denoting the number of type $I$ and type $II$ lesions, respectively. The heuristic interpretation of the coefficients in equation \eqref{EQN:LQM} is that $a$ is the rate at which a lesion of type $II$ becomes a lesion of type $I$, $r$ is the rate at which a lesion of type $II$ recovers and goes to the set $\emptyset$ (i.e. that of the healthy cells), whereas $b$ is the rate at which two lesions interact to become a single type $I$ lesion. These considerations can be mathematically expressed as
\begin{equation}\label{EQN:React}
\begin{split}
& X \xrightarrow{a} Y\,,\\
& X \xrightarrow{r} \emptyset\,,\\
& X + X \xrightarrow{b} Y\,.\\
\end{split}
\end{equation}

Thus, at a given time  $t$, the probability to observe $x$ lesions of type $II$ and $y$ of type $I$ is
\[
p(t,y,x) = \mathbb{P}\left (\left (Y(t),X(t)\right ) = \left (y,x\right )\right )\,.
\] 
Also, 
\[
\begin{split}
&p_{t_0,y_0,x_0}(t,y,x) := p(t,y,x|t_0,y_0,x_0) =\mathbb{P}\left (\left .\left (Y(t),X(t)\right ) = \left (y,x\right )\right |\left (Y(t_0),X(t_0)\right ) = \left (y_0,x_0\right )\right )\\
\end{split}
\] 
is the probability conditioned to the fact that at $t=t_0$ there were $x_0$ and $y_0$ sub--lethal and lethal lesions, respectively.

To determine the governing master equation for the above probability density $p(t,y,x)$, we need to account for all possible system changes in the infinitesimal time interval $dt$

Thus, the following scenarios may happen:
\begin{description}\label{DES:react}
\item[(i)] at time $t$ we have exactly $(y,x)$ lesions and they remain equal with a rate $(1- (a+r) x - b x(x-1))dt$, namely

\[
\begin{split}
&\mathbb{P}\left (\left .\left (Y(t+dt),X(t+dt)\right ) = \left (y,x\right )\right |\left (Y(t),X(t)\right ) = \left (y,x\right )\right ) =\\
&= 1- ((a+r) x - b x(x-1))dt + O(dt^2)\,;
\end{split}
\]

\item[(ii)] at time $t$ we have exactly $(y,x+1)$ lesions, and one lesion recovers with rate $(x+1)r dt$, namely

\[
\begin{split}
&\mathbb{P}\left (\left .\left (Y(t+dt),X(t+dt)\right ) = \left (y,x\right )\right |\left (Y(t),X(t)\right ) = \left (y,x+1\right )\right )= (x+1)r dt + O(dt^2)\,;
\end{split}
\]

\item[(iii)] at time $t$ we have exactly $(y-1,x+1)$ lesions, and one type $II$ lesion becomes of type $I$ with a rate $(x+1) a dt$, namely

\[
\begin{split}
&\mathbb{P}\left (\left .\left (Y(t+dt),X(t+dt)\right ) = \left (y,x\right )\right |\left (Y(t),X(t)\right ) = \left (y-1,x+1\right )\right ) = (x+1) a dt + O(dt^2)\,;
\end{split}
\]

\item[(iv)] at time $t$ we have exactly $(y-1,x+2)$ lesions, and two type $II$ lesions become one type $I$ with a rate $(x+2)(x+1)b dt$, namely

\[
\begin{split}
&\mathbb{P}\left (\left .\left (Y(t+dt),X(t+dt)\right ) = \left (y,x\right )\right |\left (Y(t),X(t)\right ) = \left (y-1,x+2\right )\right ) = (x+2)(x+1)b dt + O(dt^2)\,;
\end{split}
\]

\end{description}  

Grouping the equations derived in \ref{DES:react} we obtain
\[
\begin{split}
p(t+dt,y,x) &=p(t,y,x)\left (1- ((a+r) x - b x(x-1))dt + O(dt^2)\right ) + \\
&+p(t,y,x+1)\left ((x+1)r dt + O(dt^2)\right ) + \\
&+p(t,y-1,x+1)\left ((x+1) a dt + O(dt^2)\right ) +\\
&+ p(t,y-1,x+2)\left ((x+2)(x+1)b dt + O(dt^2)\right )\,,
\end{split}
\]

Writing down above relation and taking the limit as $dt \to 0$ we eventually obtain the \textit{microdosimetric master equation} (MME)
\begin{equation}\label{EQN:Master}
\begin{split}
\partial_t p(t,y,x) &= - \left ((a+r) x - b x(x-1)\right )p(t,y,x)+ (x+1)r p(t,y,x+1) + \\
&+(x+1)a p(t,y-1,x+1) +(x+2)(x+1)b p(t,y-1,x+2)\,,
\end{split}
\end{equation}
where above $\partial_t$ denotes the partial derivative with respect to the first argument of $p(t,y,x)$, that is the time variable. Equation \eqref{EQN:Master} must be equipped with suitable initial condition $p(0,y,x) = p_0(y,x)$.

We remark that the above derived MME arises solely from the probabilistic assumptions regarding lesion formation.

The MME \eqref{EQN:Master} can be written for short as
\begin{equation}\label{EQN:Master2}
\begin{split}
\partial_t p(t,y,x) &= \left (E^{-1,2} -1\right )\left [x(x-1) b p(t,y,x)\right ] + \left (E^{-1,1} -1\right )\left [x a p(t,y,x)\right ]+\left (E^{0,1} -1\right )\left [x r p(t,y,x)\right ]=\\
&= \EE^{-1,2}\left [x(x-1) b p(t,y,x)\right ] + \EE^{-1,1} \left [x a p(t,y,x)\right ]+ \EE^{0,1} \left [x r p(t,y,x)\right ]\,,\\
\end{split}
\end{equation}
where above we have denoted the creation operators defined as
\[
\begin{split}
\EE^{i,j} \left [f(t,y,x)\right ] &:= \left (E^{i,j}-1\right )\left  [f(t,y,x)\right  ] :=f(t,y+i,x+j) - f(t,y,x)\,.
\end{split}
\]

\subsection{Connection with the MKM}\label{SEC:Conn}

The present section aims at showing that the mean value of the master equation does satisfy, under certain assumptions, the kinetic equations \eqref{EQN:LQM}. In what follows, $\mathbb{E}$ denotes the mean value of a random variable defined as
\[
\begin{split}
\bar{x}(t) &:= \mathbb{E}[X(t)] = \sum_{x,y \geq 0} x p(t,y,x)\,,\\
\bar{y}(t) &:= \mathbb{E}[Y(t)] = \sum_{x,y \geq 0} y p(t,y,x)\,.
\end{split}
\]

Note that, for a general function $f$, the following holds true
\begin{equation}\label{EQN:CreaOp}
\begin{split}
\sum_{x,y \geq 0} x \EE^{i,j} \left [f(y,x) p(t,y,x)\right ] &= -\mathbb{E} j f(Y,X)\,, \\
\sum_{x,y \geq 0} y \EE^{i,j} \left [f(y,x) p(t,y,x)\right ] &= -\mathbb{E} i f(Y,X)\,.
\end{split}
\end{equation}

Therefore, multiplying the MME \eqref{EQN:Master2} by $x$ and $y$, we obtain using \eqref{EQN:CreaOp}
\begin{equation}\label{EQN:MeanM}
\begin{cases}
\frac{d}{dt}\mathbb{E}[Y(t)] &= b \mathbb{E}[X(t)(X(t)-1)] + a \mathbb{E}[X(t)] \,,\\
\frac{d}{dt}\mathbb{E}[X(t)] &= - 2 \mathbb{E}[X(t)(X(t)-1)] - (a+r) \mathbb{E}[X(t)] \,.\\
\end{cases}
\end{equation}

Equations \eqref{EQN:MeanM} are still not of the form of equations \eqref{EQN:LQM2}; in particular they depend on a second order moment $\mathbb{E}[X(t)(X(t)-1)]$. Nonetheless explicit computation will show that, if we try to compute a kinetic equation for the second order moment $\mathbb{E}[X(t)(X(t)-1)]$, we would obtain a dependence on higher moments, and so to obtain an infinite set on coupled ODE. To solve the impasse we shall make what is called a \textit{mean--field} assumption, that is we assume that 
\[
\mathbb{E}[X(t)(X(t)-1)] \sim \mathbb{E}[X(t)]^2\,.
\]

Under the above \textit{mean--field assumption}, equations \eqref{EQN:MeanM} become
\begin{equation}\label{EQN:MeanM2}
\begin{cases}
\frac{d}{dt}\bar{y}(t) &= b \bar{x}^2(t) + a \bar{x}(t) \,,\\
\frac{d}{dt}\bar{x}(t) &= - 2 \bar{x}^2(t) - (a+r) \bar{x}(t) \,,\\
\end{cases}
\end{equation}
and the original kinetic equations are in turn recovered.

A quick remark on the \textit{mean--field assumption} is needed. In the case of $x$ being large enough, we have that the following approximation holds true $\mathbb{E}[X(t)(X(t)-1)] \sim \mathbb{E}[X^2(t)]$; therefore the \textit{mean field assumption} means that $\mathbb{E}[X^2(t)] - \mathbb{E}[X(t)]^2 \sim 0$. Noticing that the last term is nothing but the variance, and recalling that the variance for a random variable is null if and only if the random variable is in fact deterministic, if the \textit{mean field assumption} is realistic than the realized number of lesion does not differ much from the mean value so that everything we need to know is the mean value.
On the contrary if there are evidence that the mean value is not a realistic approximation for the realized number of lesion, the \textit{mean--field assumption} must be considered unrealistic so that the knowledge of the full probability distribution is essential to have a complete understanding of the system. 
\subsection{On the initial distribution for the number of lethal and sub-lethal lesions}\label{SEC:InitD}

One of the main advantages of the proposed model is that the distribution of DNA damages induced by an ionizing radiation $z$ does not need to be chosen as Poissonian. In the present section we will show how the number of induced lesions can be evaluated starting from microdosimetric spectra. 

Let $f_{1;d}(z)$ be the single--event distribution of energy deposition on a domain $d$, see \cite[]{Ros}. The single--event distribution $f_{1;d}$ can be either computed numerical via Monte Carlo toolkit or by experimental microdosimetric measurements. 

The full probability distribution of an energy deposition thus depends on the number of events that deposit energy on the cell nucleus. Given a cell nucleus, composed by $N_d$ domains, the probability that $\nu$ events deposit an energy $z$ obeys to a Poissonian distribution of mean $\lambda_n := \frac{z_n}{z_F}$, being $z_n$ the mean energy deposition on the nucleus, i.e.
\[
z_n = \int_0^\infty z f(z|z_n) dz\,,
\]
and $z_F$ the first moment of the single event distribution $f_{1;d}$ defined as
\begin{equation}\label{EQN:ZF}
z_F := \int_0^\infty z f_{1;d}(z) dz\,.   
\end{equation}

Then, assuming a Poissonian probability that a domain registers $\nu$ events, the energy deposition distribution is given by
\[
f(z|z_n) := \sum_{\nu = 0}^\infty \frac{e^{- \frac{z_n}{z_F}}}{\nu!}\left (\frac{z_n}{z_F}\right )^n f_{\nu;d}(z)\,,
\]
where $f_{\nu;d}(z)$ is the energy deposition distribution resulting from $\nu$ depositions.

In particular, given a domain $d$ suffers $\nu$ energy deposition events, the distribution resulting from $\nu$ events can be computed convolving $\nu$ times the single event distribution, see, \cite{Ros,Sat}. Therefore, the imparted energy $z$ has distribution $f_{\nu;d}$, computed iteratively as
\[
\begin{split}
f_{2;d}(z) &:= \int_0^\infty f_{1;d}(\bar{z})f_{1;d}(z-\bar{z})d\bar{z}\,,\\
&\dots\,,\\
f_{\nu;d}(z) &:= \int_0^\infty f_{1;d}(\bar{z})f_{\nu-1;d}(z-\bar{z})d\bar{z}\,.\\
\end{split}
\]

For a certain energy deposition $z$, the induced number of lesions is a random variable. The standard assumption is that the distribution of $X$ given $z$ is a Poisson random variable of mean value $\kappa z$. Analogous reasoning holds for $Y$, being the number of induced lesion given $z$ a Poisson random variable of mean $\lambda z$. Given the high--flexibility of the proposed approach, the number of induced lesions given an energy deposition $z$ can be any random variables. It is worth stressing that the chosen distribution may vary with LET.

In the following general treatment we will denote by $p^X_z(x|\kappa z)$, resp. $p^Y_z(y|\lambda z)$, the initial random distribution for the number of sub--lethal, resp. lethal, lesions given an energy deposition $z$. We remark again that both $p^X_z(x|\kappa z)$ and $p^Y_z(y|\lambda z)$ can be any probability distribution. Specific relevant examples will be considered in the numerical implementation.

Putting all the above reasoning together, the MME \eqref{EQN:Master2} reads
\begin{equation}\label{EQN:MMEInitial}
\begin{cases}
\partial_t p(t,y,x) &= \EE^{-1,2}\left [x(x-1) b p(t,y,x)\right ] + \EE^{-1,1} \left [x a p(t,y,x)\right ]+ \EE^{0,1} \left [x r p(t,y,x)\right ]\,,\\
p(0,y,x) &= p^X_0(x)p^Y_0(y)\,,
\end{cases}
\end{equation}
where the initial distribution is obtained as
\begin{equation}\label{EQN:InitialIntegral}
\begin{split}
p^X_0(x) &= \int_0^\infty p^X_z(x|\kappa z) f(z|z_n) dz \,,\\
p^Y_0(y) &= \int_0^\infty p^Y_z(y|\kappa z) f(z|z_n) dz\,.\\
\end{split}
\end{equation}

\subsection{The protracted dose case for the Generalized Stochastic Microdosimetric Model}\label{SEC:Split}

The MME can be further generalized to consider \textit{protracted dose} irradiation. We refer to \textit{protracted dose} as a continuous dose delivery in time. On the contrary a fixed in time, asymptotically short impulse-like dose irradiation is called \textit{acute dose irradiation}, whereas a series of acute irradiations  at prescribed timesteps is referred to as \textit{split dose irradiation}. Existing models fail at properly describing protracted dose, being unable to fully capture the stochasticity inherent to energy deposition. Usually, strong assumptions are used to treat protracted dose, \cite{Haw2}, or a split dose is used to approximate a continuous dose delivery, \cite{Ina}. Nonetheless, models cannot fully predict experimental data, \cite{Ina}.

The generalization of the GSM$^2$ master equation \eqref{EQN:Master2} to consider a continuous dose irradiation is not trivial. In fact, at random time $t$ the number of lesions, either lethal or sublethal, exhibits a jump upward of a random quantity that depends on the energy deposition $z$, that we recall is a random variable.

More formally, the possible interactions now become
\[
\begin{split}
& X \xrightarrow{a} Y\,,\\
& X \xrightarrow{r} \emptyset\,,\\
& X + X \xrightarrow{b} Y\,.\\
& X \xrightarrow{\dot{d}} X + Z_{\kappa}\,.\\
& Y \xrightarrow{\dot{d}} Y + Z_{\lambda}\,,\\
\end{split}
\]
being $Z_{\lambda}$ and $Z_{\kappa}$ two random variables with integer--valued distributions $p_0^X$ and $p_0^Y$ respectively, defined as in equation \eqref{EQN:MMEInitial}. The parameter $\dot{d}$ represents the dose rate, see, \cite{HawIna,HawIna2}, and it is given by $\dot{d} := \frac{z_n}{T_{irr} z_F}$, being $z_F$ given in equation \eqref{EQN:ZF} and $T_{irr}$ is the total irradiation time.

\begin{description}\label{DES:react2}
\item[(i)]

\[
\begin{split}
&\mathbb{P}\left (\left .\left (Y(t+dt),X(t+dt)\right ) = \left (y,x\right )\right |\left (Y(t),X(t)\right ) = \left (y,x\right )\right )=\\
&= 1 - ((a+r) x + b x(x-1) + \dot{d} (1-p_0^X(0))(1-p_0^Y(0)))dt + O(dt^2)\,;
\end{split}
\]

\item[(ii)] 

\[
\begin{split}
&\mathbb{P}\left (\left .\left (Y(t+dt),X(t+dt)\right ) = \left (y,x\right )\right |\left (Y(t),X(t)\right ) = \left (y-i_y,x-i_x\right )\right ) =\\
&= \dot{d} p_0^X(i_x) p_0^Y(i_y) dt + O(dt^2)\,, \quad i_x = 1,\,\dots,x\,,i_y = 1,\,\dots,y\,,
\end{split}
\]
\end{description}  

Further, reactions $(ii)$, $(iii)$ and $(iv)$ in Section \ref{SEC:ME} remain valid.

Therefore, a similar analysis to the one carried out in Section \ref{SEC:ME} leads to the following MME
\begin{equation}\label{EQN:Master2FullSplit}
\begin{split}
\partial_t p(t,y,x) &= \left (E^{-1,2} -1\right )\left [x(x-1) b p(t,y,x)\right ] + \left (E^{-1,1} -1\right )\left [x a p(t,y,x)\right ]+\\
&+\left (E^{0,1} -1\right )\left [x r p(t,y,x)\right ]+\left (\sum_{i_x=1}^x \sum_{i_y=1}^y E^{-i_y,-i_x}_{\dot{d}} - (1-p_0^X(0)) (1-p_0^Y(0))\right )\left [\dot{d} p(t,y,x)\right ]= \\
&= \EE^{-1,2}\left [x(x-1) b p(t,y,x)\right ] + \EE^{-1,1} \left [x a p(t,y,x)\right ]+ \\
&+\EE^{0,1} \left [x r p(t,y,x)\right ]+\EE^{-y,-x}_{\dot{d}}\left [ \dot{d} p(t,y,x)\right ] \,.
\end{split}
\end{equation}

The operator in the last line of equation \eqref{EQN:Master2FullSplit} right end side has been defined as

\[
\begin{split}
&\EE^{-y,-x}_{\dot{d}} f(t,y,x) := \left (\sum_{i_x=1}^x \sum_{i_y=1}^y E^{-i_y,i_x}_{\dot{d}} - (1-p_0^X(0))(1-p_0^Y(0))\right )f(t,y,x) =\\
&=\sum_{i_x=1}^x \sum_{i_y=1}^yp_0^X(i_x) p_0^Y(i_y)\, f(t,y-i_y,x-i_x) - (1-p_0^X(0))(1-p_0^Y(0))\, f(t,y,x)\,.
\end{split}
\]

The protracted dose is assumed to be delivered up to a finite time $T_{irr}<\infty$, beyond which no irradiation is considered and the systems evolves according to \eqref{EQN:Master2}.

\subsection{The diffusive cell nucleus model for GSM$^2$}\label{SEC:Vox}

In Section \ref{SEC:ME}, we investigated the time evolution for lethal and sub--lethal lesions in the cell nucleus. 

As we discussed above, one of the major weaknesses of the standard MKM and its extensions is the choice of the cell domains \cite{Smi}. In fact, too small domains translate in a null probability of double events, whereas too big domains imply that distant lesions may combine to produce a lethal lesion. To overcome this problem, the cell nucleus is split into several domains so that the time evolution in each domain can be considered independently. Further, following treatment's aim is to encompass above limitations, allowing domains interaction and variability in shape and dimension.

In the current Section, we will show how the MME \eqref{EQN:Master2FullSplit} can be extended to include interactions between the domains. In order to keep the treatment as clear as possible, no protracted dose will be consider. The general case of a continuous irradiation can easily included in the following treatment via arguments analogous to the ones used in Section \ref{SEC:Split}.

Let us consider $N_d$ domains (referred to also as voxels) that can undergo one of the following possible reactions
\begin{equation}\label{EQN:ReactN}
\begin{split}
& X_i \xrightarrow{a} Y_i\,,\quad i=1,\dots,N_d\,,\\
& X_i \xrightarrow{r} \emptyset\,,\quad i=1,\dots,N_d\,,\\
& X_i + X_i \xrightarrow{b} Y_i\,,\quad i=1,\dots,N_d\,.
\end{split}
\end{equation}

A reasoning analogous to the one carried out in Section \ref{SEC:ME} leads to the following MME
\begin{equation}\label{EQN:MasteriN}
\begin{split}
\partial_t p(t,y,x) &= \sum_{i=1}^{N_d} \EE^{-1,2}_i\left [x_i(x_i-1) b p(t,y,x)\right ] + \sum_{i=1}^{N_d} \left ( \EE^{-1,1}_i \left [x_i a p(t,y,x)\right ]+ \EE^{0,1}_i\left [x_i r p(t,y,x)\right ]\right ) \,.
\end{split}
\end{equation}
In equation \eqref{EQN:MasteriN}, the variables $x$ and $y$ are $N-$ dimensional vectors with $i-$th component given by $x_i$ and $y_i$, representing the number of sub--lethal or lethal lesions, respectively, within the $i-$th domain ($i =1,\,\dots, N$).

\begin{Remark}
In order to keep the notation as simple as possible, in equation \eqref{EQN:ReactN} we chose the rates $a$, $b$ and $r$ independent of the domain. Similar results would be obtained with voxel-dependent rates $a_i$, $b_i$ and $r_i$, $i=1,\,\dots,\,N_d$.
\end{Remark}

Empirical evidence shows that the lesions, together with interacting within the same voxel, may also move to a different voxel. In fact, lesion spatial movements inside a cell has been demonstrated to be significantly higher than the typical voxel size \cite{Schet} . To account for this behaviour, we will add an additional term to the MME \eqref{EQN:MasteriN}.

Besides reactions considered in equation \eqref{EQN:ReactN}, we now assume further the following 
\begin{equation}\label{EQN:ReactV}
\begin{split}
& X_i \xrightarrow{\kappa_{i;j}^X} X_j\,, \quad i,\,j=1,\dots,N_d\,,\\
& Y_i \xrightarrow{\kappa_{i;j}^Y} Y_j\,, \quad i,\,j=1,\dots,N_d\,.\\
\end{split}
\end{equation}

\begin{Remark}
We assumed possible interactions also between non adjacent domains. If the reactions described by equation \eqref{EQN:ReactV} are to be intended as lesions movements inside the cell nucleus, the most reasonable choice for the interaction rates is to set
\[
\kappa_{i;j}^X = \kappa_{i;j}^Y = 0\,,
\]
for $j \not \in \Gamma_i$, being $\Gamma_i$ the set of adjacent domains to $i$.
\end{Remark}

Following the same process described in in Section \ref{SEC:ME}, we obtain the MME
\begin{equation}\label{EQN:MasteriND}
\begin{split}
\partial_t p(t,y,x) &= \sum_{i=1}^N \EE^{-1,2}_i\left [x_i(x_i-1) b p(t,y,x)\right ] + \sum_{i=1}^N\left ( \EE^{-1,1}_i \left [x_i a p(t,y,x)\right ]+ \EE^{0,1}_i\left [x_i r p(t,y,x)\right ]\right )+\\
&+ \sum_{i,j=1}^N {}^X\EE^{-1,1}_{i,j} \left [x_i \kappa_{i;j}^X p(t,y,x)\right ] +\sum_{i,j=1}^N {}^Y\EE^{Y;-1,1}_{i,j} \left [y_i \kappa_{i;j}^Y p(t,y,x)\right ]\,,
\end{split}
\end{equation}
where the operators are defined as
\[
\begin{split}
{}^X\EE^{-1,1}_{i,j} f(t,y,x) &= \left (E^{0,1}_{i}E^{0,-1}_{j} -1\right )f(t,y,x) \,,\\
{}^Y\EE^{-1,1}_{i,j} f(t,y,x) &= \left (E^{1,0}_{i}E^{-1,0}_{j} -1\right )f(t,y,x) \,.\\
\end{split}
\]

The first two lines of equation \eqref{EQN:MasteriND} accounts for reactions within the same voxel, whereas the last line described movements between adjacent domains.

Using the same approach for modeling the initial damage distribution (Section \ref{SEC:InitD}) the resulting MME reads
\begin{equation}\label{EQN:MasteriNDDoubleInit}
\begin{cases}
\partial_t p(t,y,x) &= \sum_{i=1}^{N_d} \EE^{-1,2}_i\left [x_i(x_i-1) b p(t,y,x)\right ] + \sum_{i=1}^{N_d}\left ( \EE^{-1,1}_i \left [x_i a p(t,y,x)\right ]+ \EE^{0,1}_i\left [x_i r p(t,y,x)\right ]\right )+\\
&+ \sum_{i,j=1}^{N_d} {}^X\EE^{-1,1}_{i,j} \left [x_i \kappa_{i;j}^X p(t,y,x)\right ] +\sum_{i,j=1}^{N_d} {}^Y\EE^{Y;-1,1}_{i,j} \left [y_i \kappa_{i;j}^Y p(t,y,x)\right ]\,,\\
p(0,y,x) &= \prod_{i=1}^N p_{0;i}^X(x_i)p_{0;i}^Y(y_i)\,,
\end{cases}
\end{equation}
where $p_{0;i}^X(x_i)p_{0;i}^Y(y_i)$ denotes the initial distribution for the voxel $i$ as computed in equations \eqref{EQN:MMEInitial}--\eqref{EQN:InitialIntegral}.

\subsection{Survival probability}\label{SEC:Surv}

Cell survival is one of the most relevant biological endpoints in radiobiology and is defined as the probability for a cell to survive radiation exposure\red, {mostly measured by its ability to form clonogens, i.e., to retain its reproductive potential}. Taking into account assumption $5$, no lethal lesions must be present into the cell nucleus after a sufficiently large time has passed from the irradiation. An estimate of cell survival can be obtained from the solution to the MME \eqref{EQN:Master2}. In this study, we will focused on a single domain, because the calculations for the entire cell is completely analogous. 

The survival probability for the domain $d$ under the assumptions $1$-$5$ introduced above, is defined as
\begin{equation}\label{EQN:Surv}
S^Y := \mathbb{P}\left (\lim_{t \to \infty} Y(t) = 0\right )\,;
\end{equation}
in order to asses the survival probability, the limiting long--time distribution for the MME \eqref{EQN:Master2} must be studied. 

From an heuristic perspective, since the number of sub--lethal lesion can only decrease, the points $\{(y,0)\,:\, y \in \mathbb{N}_0\}$ are absorbing states. Furthermore, the system reaches an absorbing state in a finite time with probability $1$, converging towards a limiting stationary distribution. With \textit{absorbing state} we mean that once the system reaches the point $(y,0)$, it stays there and future evolutions are no longer considered. 

The high generality of the GSM$^2$model, especially because no detailed balance is satisfied or explicit conserved quantities can be obtained, makes the closed form for the limiting distribution not easily computable. For this reason, in the present work the survival probability will be computed from the corresponding master equation numerical solution as 
\[
S^Y = \lim_{t \to \infty} p(t,0,0)\,.
\]

In forthcoming developments, we will study in more detail the survival probability resulting from the proposed GSM$^2$ model and its explicit form. In general, it is worth mentioning that, besides the numerical approach, as the one used here, and the analytical approach in which the survival probability is explicitly computed, an efficient approach is to introduce suitable approximations in the driving equation so that a formal expansion of the survival probability can be computed, \cite{Gar}.

\section{Numerical implementation}\label{SEC:Num}

To calculate a numerical solution to the MME \eqref{EQN:Master2}, the following steps are performed:

\begin{enumerate}
\item We choose the number $N_d$ of domains in which the cell nucleus is divided. As GSM$^2$ does not rely on any specific assumption for the probability distribution, the domains do not need to be assumed of equal size. For each domain, the \textit{single event} energy deposition distribution $f_{1;d}(z)$ is obtained with Geant4 \cite{G4} simulations.

%
%
%

\item The number of lethal and sub-lethal lesions are sampled from the distributions $p^X_0(x)$ and $p^Y_0(x)$ as derived in equation  \eqref{EQN:InitialIntegral}. The standard assumption is that $p^X_z$, resp. $p^Y_z$, is a Poisson distribution of mean $\kappa z_d$, resp. $\lambda z_d$. Given the general setting, we will compare the results with an initial Gaussian distribution of different possible variances.

\item Given the initial number of lesions, the evolution paths are simulated via the \textit{stochastic simulation algorithms} (SSA) \cite[Chapter 13]{Wei}.

\item Steps 1-3 are repeated to obtain the Monte Carlo empirical distribution of lethal and sublethal lesions over the cell nucleus;

\item The survival probability in the single domain as well as the cell nucleus are calculated from the empirical distribution obtained in step 4;
\end{enumerate}

Previous steps can be computed independently for each domain if no interaction between domains is assumed or the paths for the whole nucleus can be estimated simultaneously, in case of a dependent-voxel model. The computational effort for the latter is substantially higher. It should be noted here that developing an efficient simulation algorithm is beyond the aim of the present work and we refer to \cite{Sim} for a review of possible simulation algorithms.



\subsection{The numerical solution}

\begin{figure}[thpb]
\centering
\includegraphics[width=.45\columnwidth]{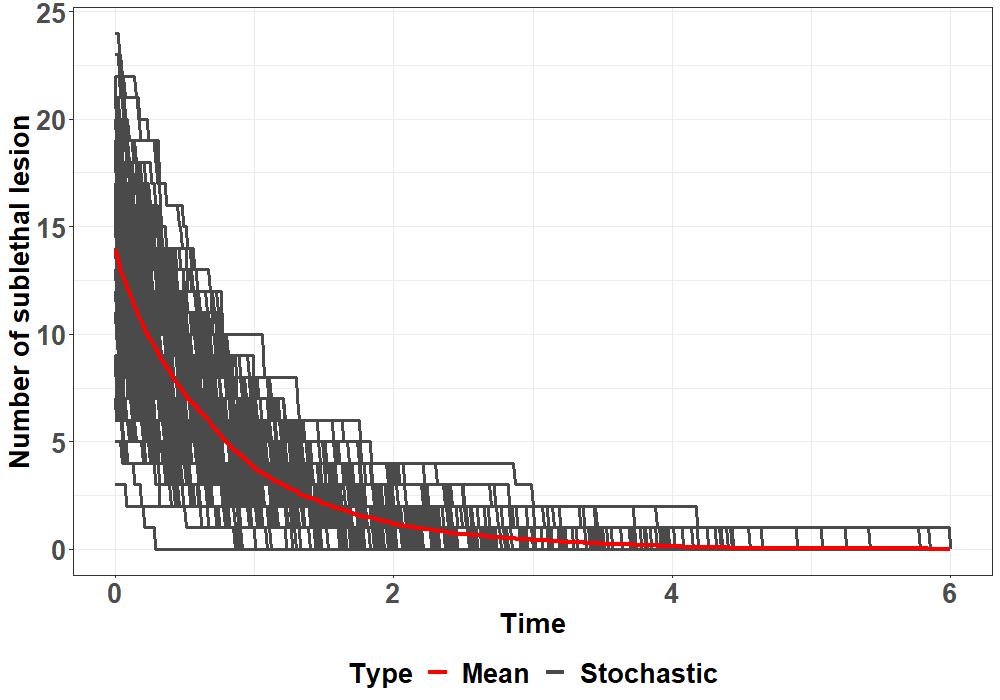}
\includegraphics[width=.45\columnwidth]{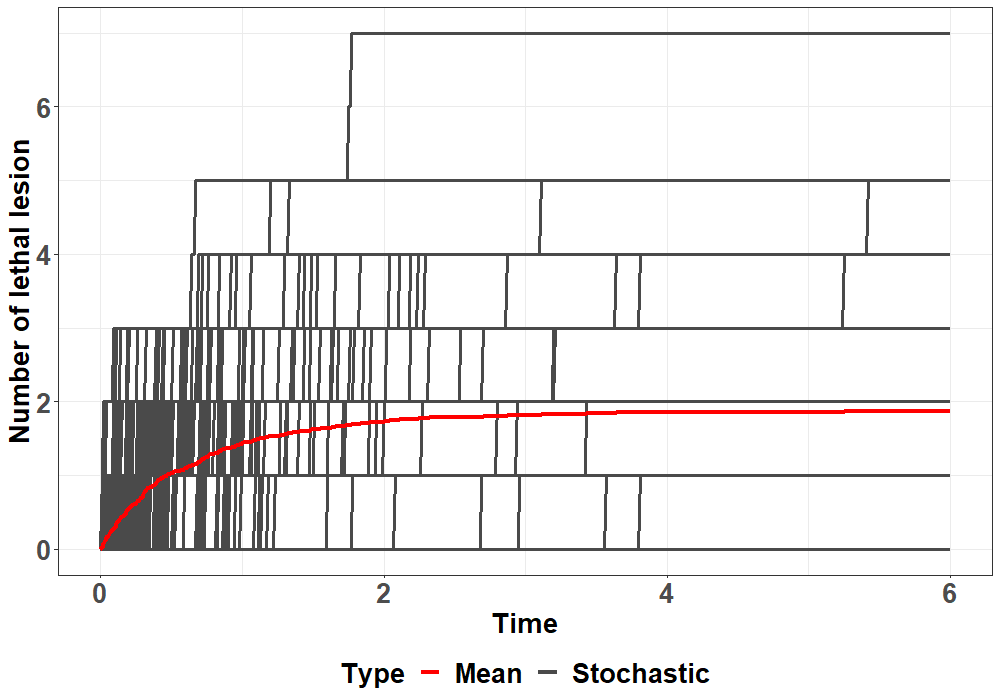}\\
\caption{Sub-lethal lesions (left panel) and lethal lesions (right panel) evolution. GSM$^2$ parameters were set to $r=1$, $a=0.1$ and $b=0.01$. The red line represents the average value.}
\label{FIG:CompM} 
\end{figure}

\begin{figure*}[!t]
\begin{tabular}{c|cc}
     \rotatebox{90}{$t1$} & \addheight{\includegraphics[width=.45\textwidth]{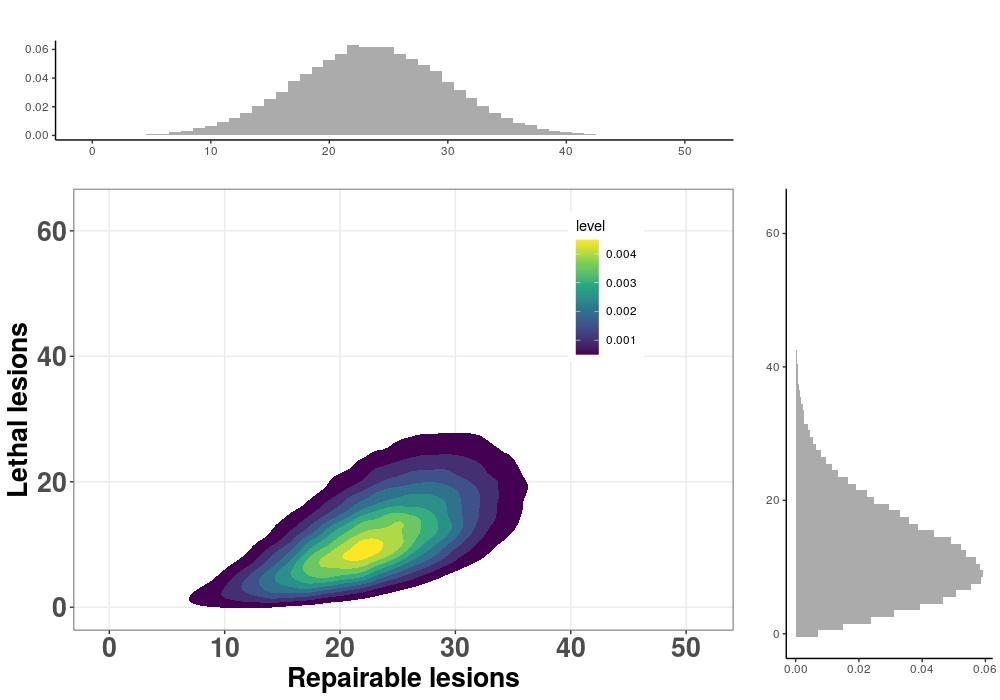}} &
      \addheight{\includegraphics[width=.45\textwidth]{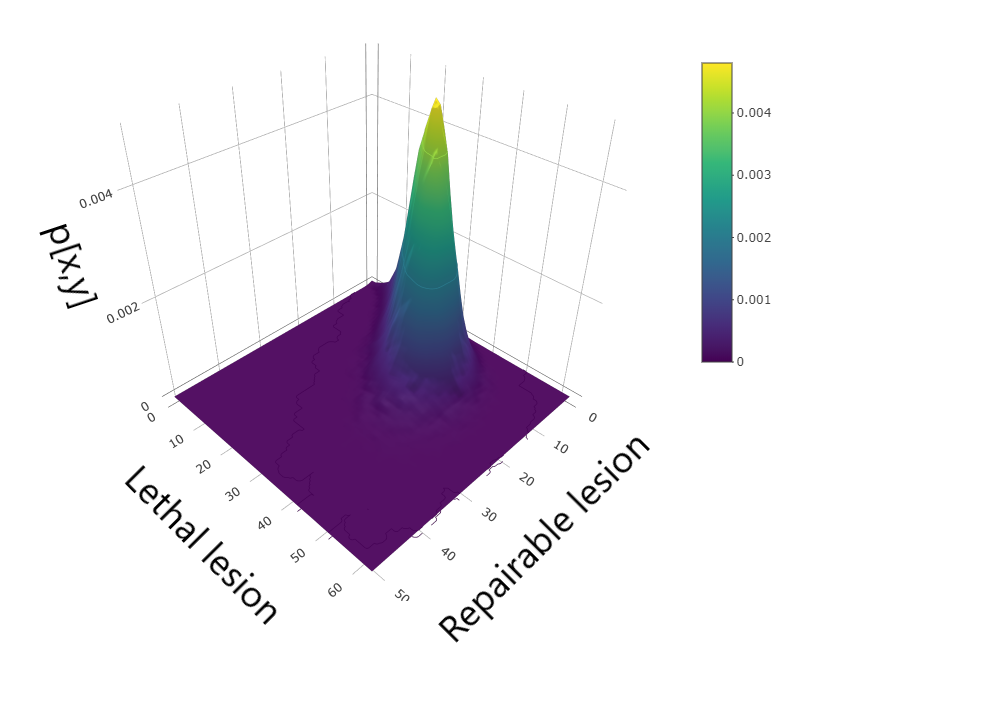}}\\
       \rotatebox{90}{$t2$} & \addheight{\includegraphics[width=.45\textwidth]{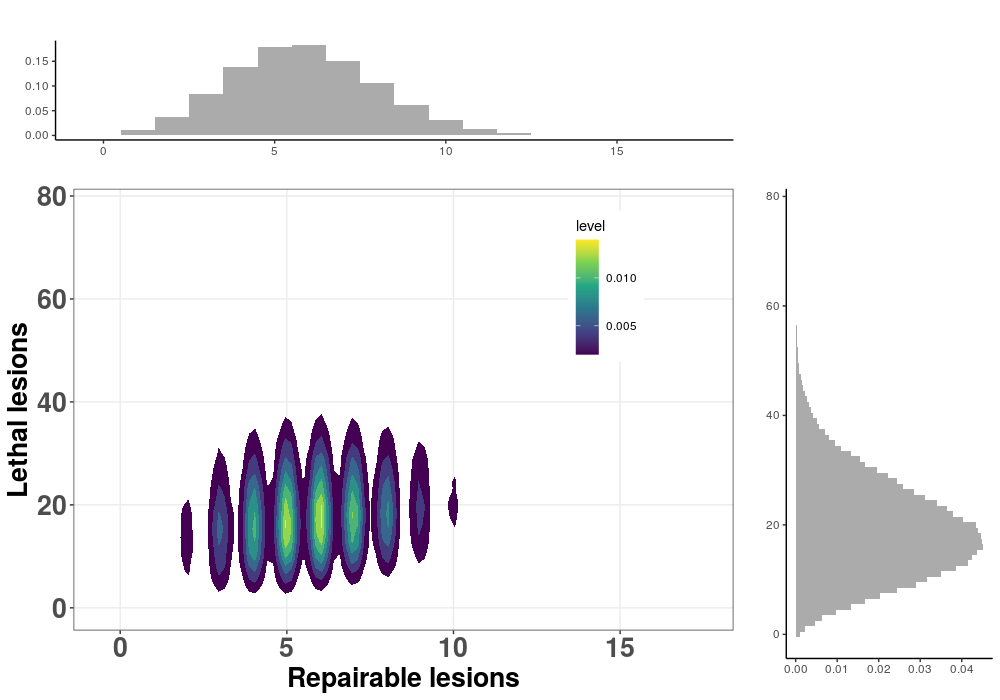}} &
      \addheight{\includegraphics[width=.45\textwidth]{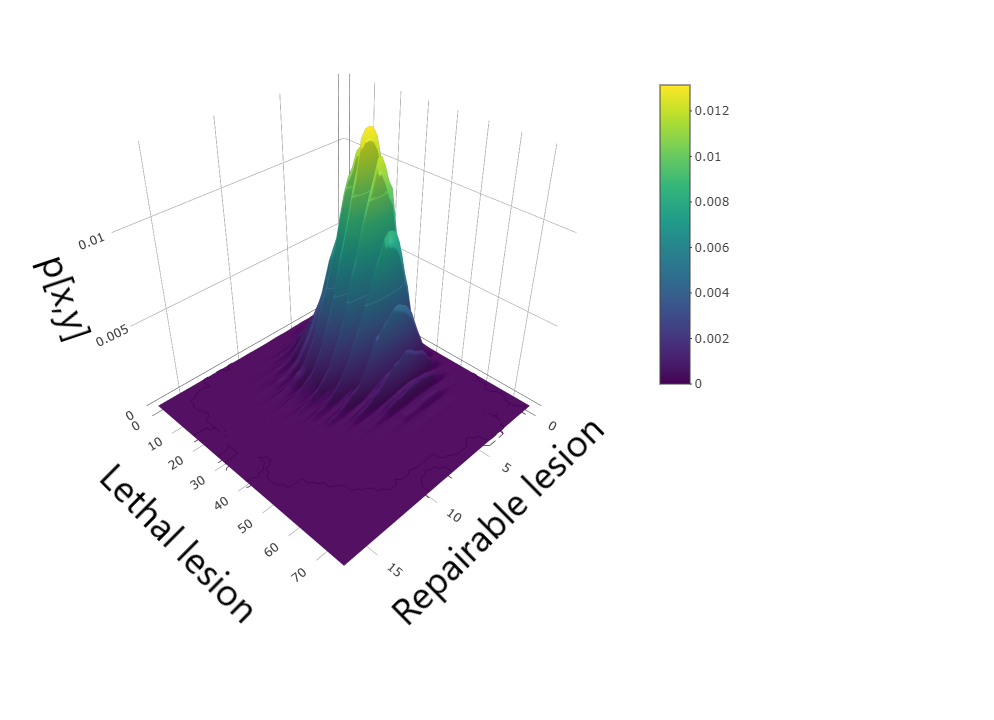}}\\
       \rotatebox{90}{$t3$} & \addheight{\includegraphics[width=.45\textwidth]{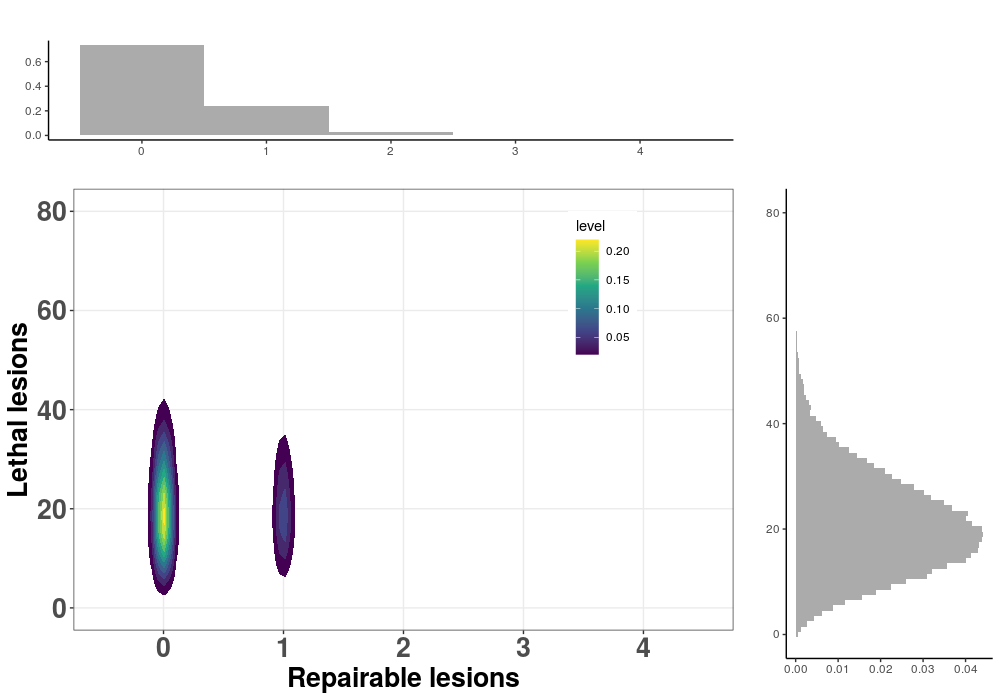}} &
      \addheight{\includegraphics[width=.45\textwidth]{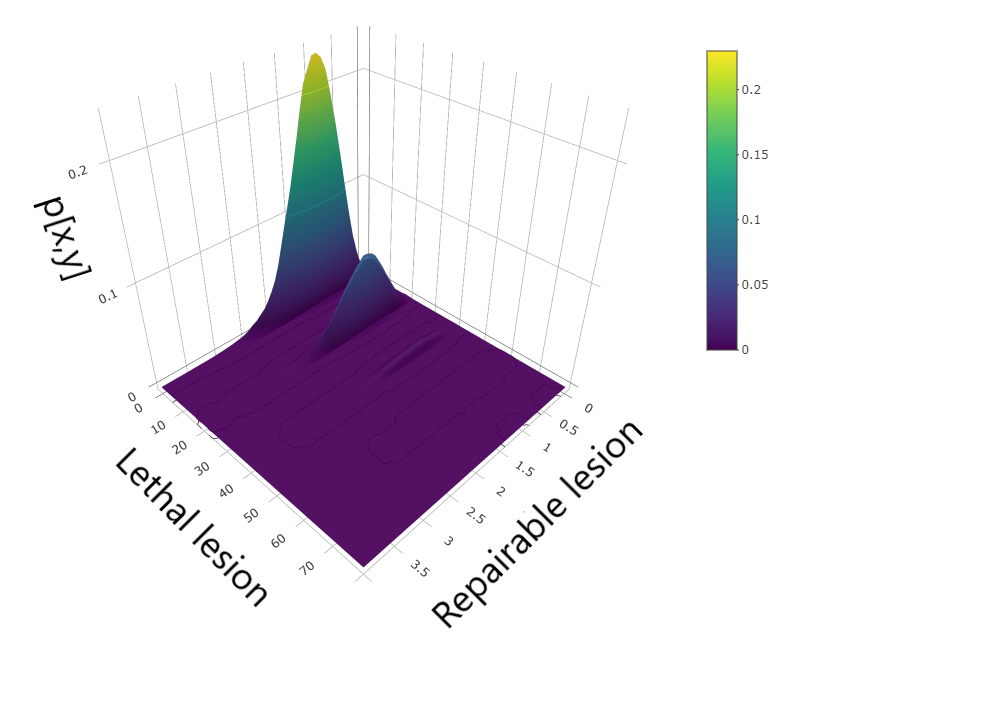}}\\
\end{tabular}
\caption{Master equation solution at time $t=1 m$ (top panel), $t=100$ arb. unit (middle panel) and $t=150$ arb. unit. GSM$^2$ parameters were set to $r=1$, $a=0.2$ and $b=0.1$.}
\label{FIG:Density} 
\end{figure*}

\begin{figure*}[!t]
\begin{tabular}{cc}
\addheight{\includegraphics[width=.45\textwidth]{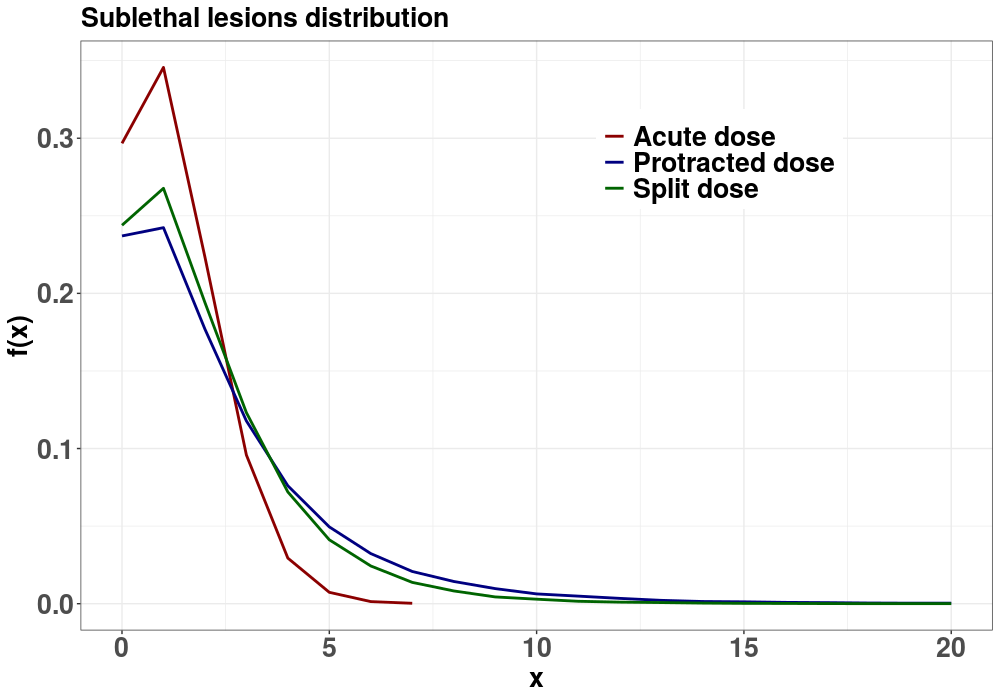}} &
      \addheight{\includegraphics[width=.45\textwidth]{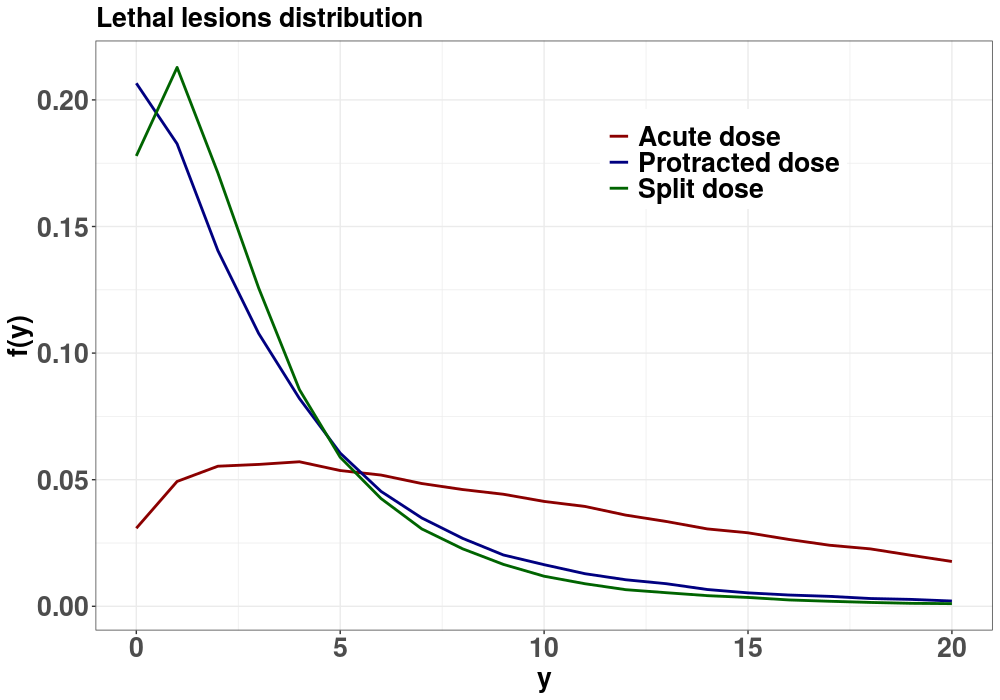}}\\
\end{tabular}
\caption{Master equation solution for acute, split and protracted doses of $100 \,Gy$. GSM$^2$ parameters were set to $r=1$, $a=0.2$ and $b=0.1$.}
\label{FIG:Dose} 
\end{figure*}

\begin{figure*}[!t]
\begin{center}
\begin{tabular}{c}
\addheight{\includegraphics[width=.65\textwidth]{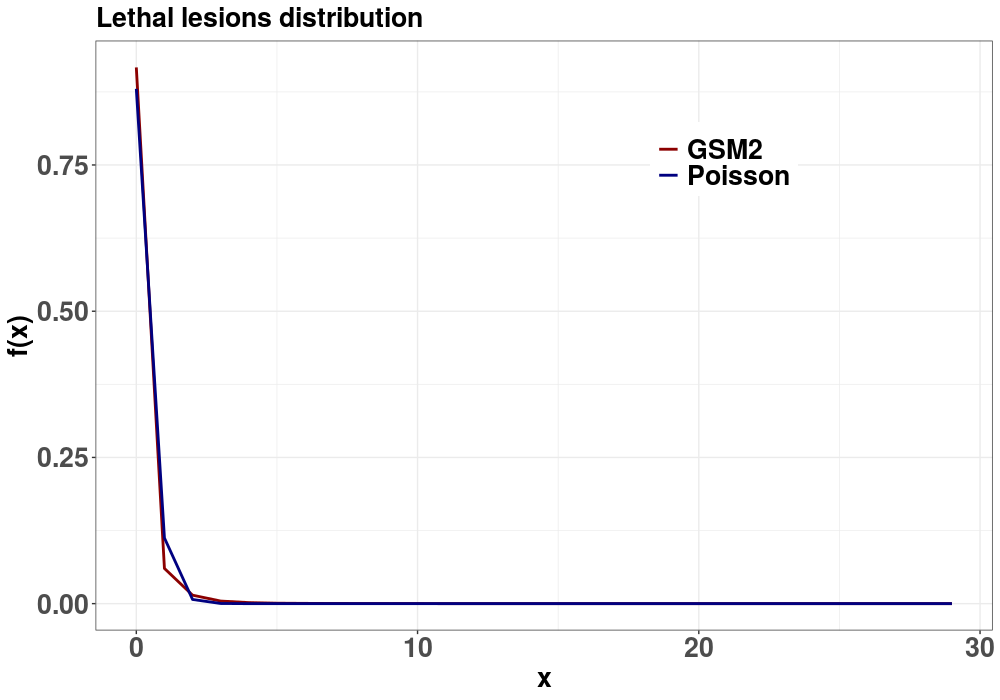}} \\
   \addheight{\includegraphics[width=.65\textwidth]{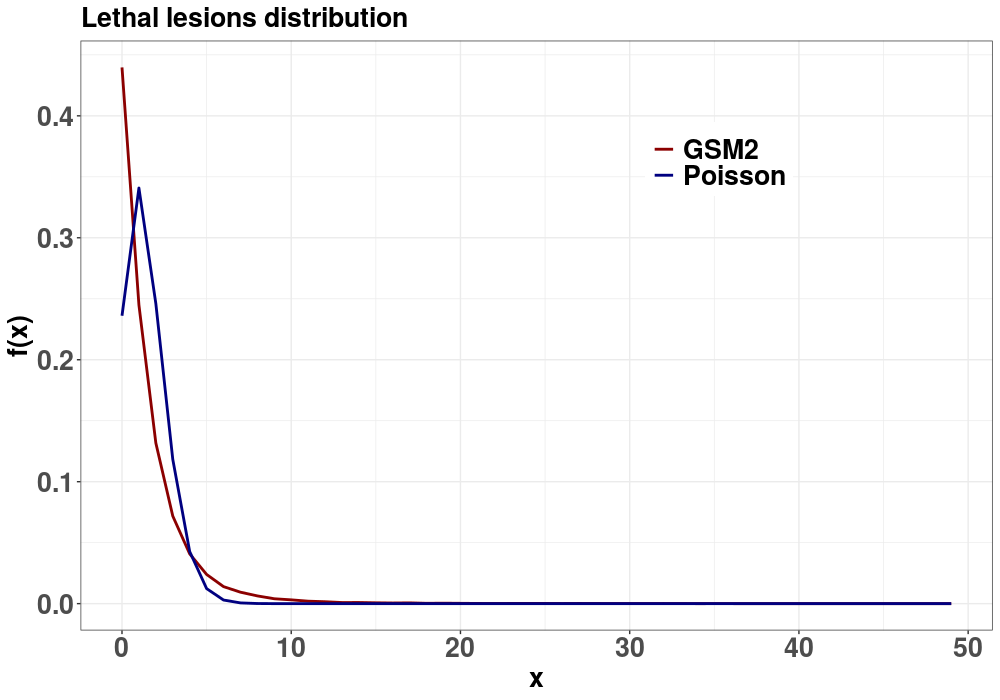}} \\
      \addheight{\includegraphics[width=.65\textwidth]{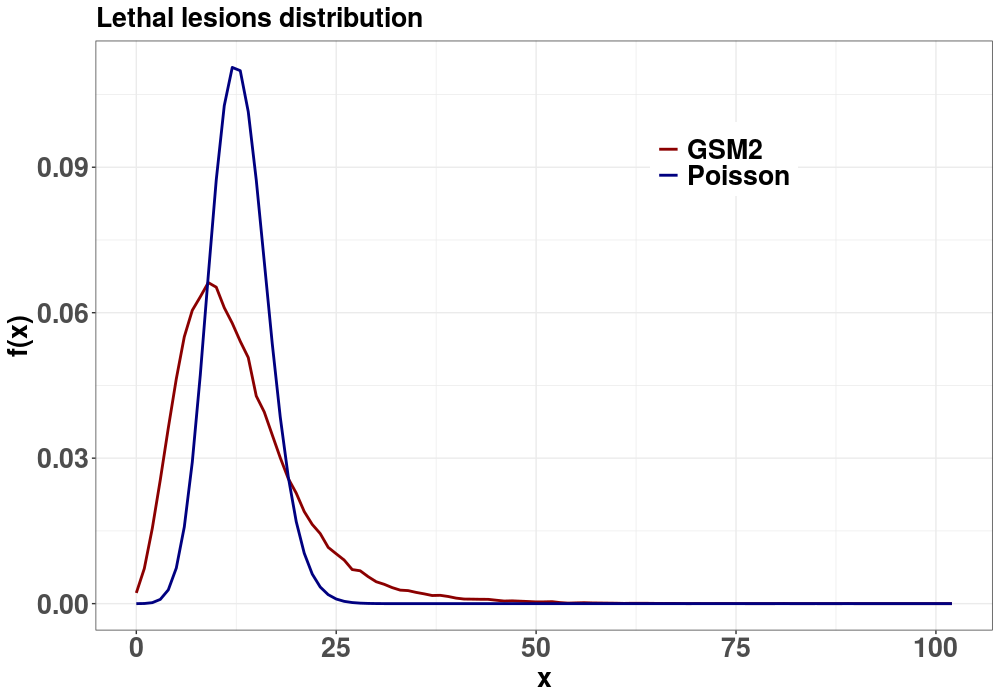}}\\
\end{tabular}
\end{center}
\caption{Comparison of long--time lethal lesion distributions and Poisson distributions. Top panel: dose=$5$ Gy, $r=1$, $a=0.1$ and $b = 0.01$. Middle panel: dose=$100$ Gy, $r=5$, $a=0.1$ and $b = 0.01$. Bottom panel: dose=$150$ Gy, $r=5$, $a=0.2$ and $b = 0.1$.}
\label{FIG:Param} 
\end{figure*}

The present Section is devoted to finding and discussing the numerical solution of MME derived in Section \ref{SEC:ME}. In particular, the full master equation \eqref{EQN:Master2} is solved via the \textit{stochastic simulation algorithms} (SSA) \cite[Chapter 13]{Wei}, so that the density is estimated with a Monte Carlo simulation. We simulate 10$^6$ events and the density function is thus reconstructed empirically. 

The goal of this Section is also to highlight how a different setting affects the lesions density distribution. In particular, it will emerge how the density distribution resulting from the corresponding master equation changes for different lesion evolution parameters, initial probabilistic conditions or also irradiation conditions. 

To assess the energy deposited on the domain, we used the microdosimetry approach as discussed in Section \ref{SEC:InitD}. With Geant4, we simulated microdosimetric spectra of a 20 MeV/u carbon ion beam traversing a 1.26 cm diameter sphere filled with pure propane gas with a low density ($1.08 *10^{-4} g/cm^3$), such that the energy depositions are equivalent to those in 2 $\mu$m of tissue. This geometry reproduces a standard Tissue Equivalent Proportional Counter (TEPC) as used for example in \cite{Mis2}. Specific energies acquired with the TEPC are then converted to the domain size of interest as reported in \cite[Section 2]{Bel}. 
The choice to simulate a microdosimeter has been made with the aim of remaining as consistent as possible with real experiments.
In addition, carbon ions have been chosen since existing model fails at predicting relevant radiobiological endpoints under high-LET regimes.

In the calculations, we consider high doses, so that multi-event distributions as described in Section \ref{SEC:InitD} are computed for $z_n >> 1$. This choice is due to the fact that the plotted distributions refer to a single cell nucleus domain and thus, to highlights differences at such a small scale, high dose needs to be considered. At lower doses, differences between the MME solution for a single nucleus domain for different parameters are more difficult to appreciate. Nonetheless, small differences at the domain level can translates into relevant dissimilarities at the macroscopic level.

Figure \ref{FIG:CompM} reports different path realizations for the lethal and sub--lethal evolution; the stochastic paths are also compared to the mean value, that evolves according to the MKM kinetic equations \eqref{EQN:LQM}. The plots indicate that the mean value can not be representative of the whole path realizations distribution.

Figure \ref{FIG:Density} shows the master equation solution at different times. The left panels show the contour plots of the joint probability distributions of lethal and sub--lethal damages, together with their marginal distributions. The rights panels are 3D representations of the density function solutions. At a starting time t$_1$, there is a high variability in the number of reparable lesions while small fluctuations are present in the number of lethal lesions. At a later time t$_3$, instead, the situation is the exact opposite, with a greater variability in the number of lethal lesions against small fluctuations in the number of sub--lethal lesions.

Figure \ref{FIG:Dose} compares lethal and sub-lethal lesion distributions for different types of irradiation conditions, namely acute dose delivery at initial time, split dose at uniform time steps and protracted dose according to equation \eqref{EQN:Master2FullSplit}. A split dose at uniform times yields a rather similar lesion distribution as a fully stochastic protracted dose irradiation, while the solution differ significantly for the acute dose case. This result is caused by the non-linear effect that double events have on the lesions probability distribution.

The long time distribution of lethal lesions is compared with a Poisson distribution for different parameters and doses in Figure \ref{FIG:Param}. At lower doses and for $b$ negligible with respect to $r$, the MME solution is in fact Poissonian (top panel).
As the dose increases, the MME solution can be non poissonian even if $r$ dominates $b$ (middle panel). Finally, for higher doses and higher $b$, the MME solution differs significantly from a Poisson distribution (bottom panel).  

\subsection{Effect of the initial law on the lethal lesions distribution and cell survival}

\begin{figure*}[!t]
\begin{tabular}{cc}
\addheight{\includegraphics[width=.45\textwidth]{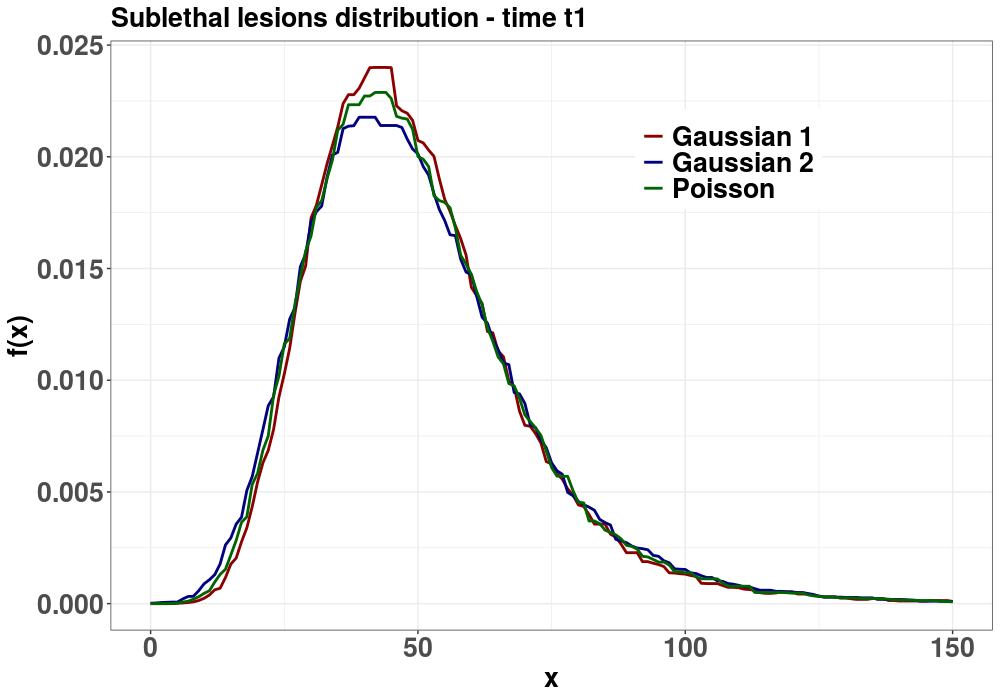}} &
      \addheight{\includegraphics[width=.45\textwidth]{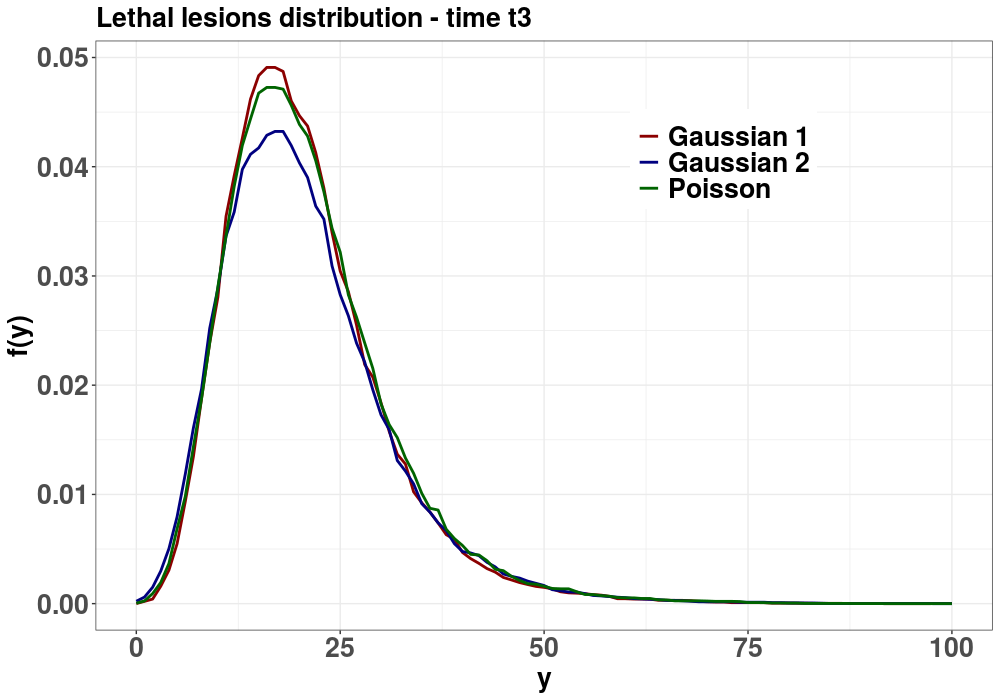}}\\
\end{tabular}
\caption{Lethal and sub-lethal lesions distribution depending on the chosen initial distribution at time $t1=1\,$ arb. unit and $t3 = 150$ arb. unit. The initial distributions $p^X_z$ and $p^Y_z$ have been chosen as a Poisson distribution of mean $\mu = {\lambda,\kappa}z$ or as a Gaussian distribution with mean $\mu = {\lambda,\kappa}z$ and variance $\sigma^2 \in \{0.5 \mu,\,,\,1.5\mu\}$. The MME parameters were set to $r=1$, $a=0.2$ and $b=0.1$.}
\label{FIG:Density2} 
\end{figure*}

\begin{figure}[thpb]
\centering
\includegraphics[width=.7\columnwidth]{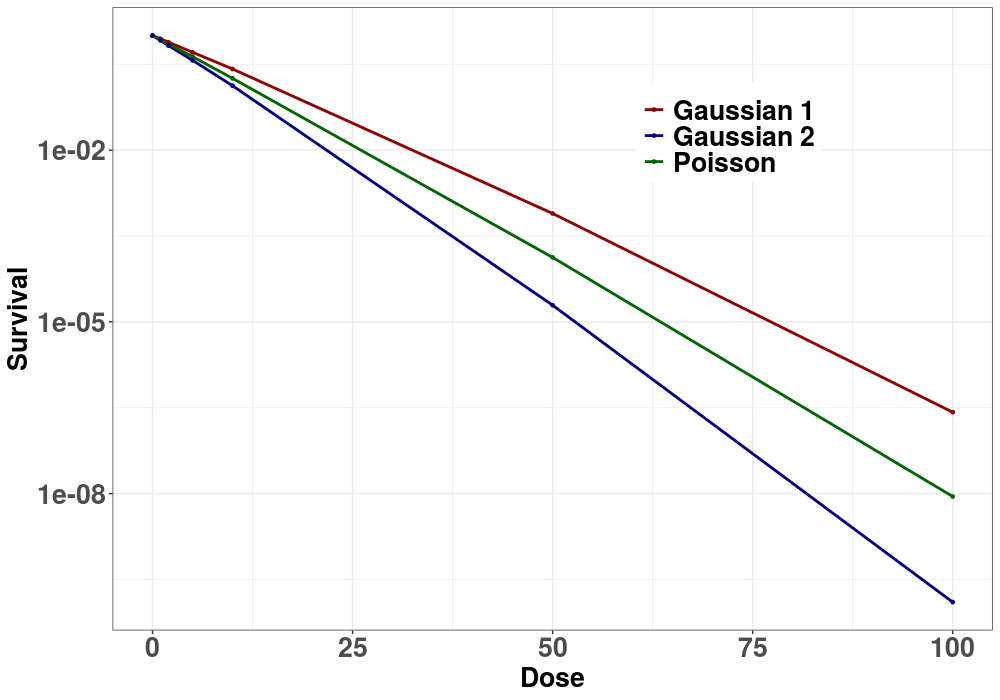}
\caption{Cell survival function calculated for different initial conditions. The initial distributions $p^X_z$ and $p^Y_z$ have been chosen as a Poisson distribution of mean $\mu = {\lambda,\kappa}z$, or as a Gaussian distribution with mean $\mu = {\lambda,\kappa}z$ and variance $\sigma^2 \in \{0.5 \mu,\,,\,1.5\mu\}$. The MME parameters were set to $r=1$, $a=0.2$ and $b=0.1$.}
\label{FIG:Survival} 
\end{figure}

The goal of the present Section is to emphasize the dependence on the initial law of the long--time lethal lesions distribution, showing that the lethal lesions marginal distribution might differ from the Poisson distribution that is typically assumed. 

We considered different initial conditions for equation \eqref{EQN:InitialIntegral}. In particular, the following initial distributions were selected for $p^X_z(x|\kappa z)$ and $p^Y_z(y|\lambda z)$: i) a Poisson random variable with mean value $\mu$; and ii) a Gaussian with mean value $\mu$ and variance between 0.5$\mu$ and 1.5$\mu$. The mean value $\mu$ has been set to $\lambda$z for sub--lethal lesions and $\kappa$z for lethal lesions.
The results are plotted in Figure \ref{FIG:Density2} and indicate that a a more peaked initial distribution correspond to a more peaked long--time distribution, meaning that the initial value can sharpen or broaden lethal and sub--lethal lesion distributions. This effect has a straightforward consequence on the resulting survival probability shown in figure \ref{FIG:Survival}.

we test both the typically used Poisson initial distribution and a Gaussian random variable with different variance.

Figure \ref{FIG:Density2} shows the comparison of lethal and sub--lethal lesion distributions for different initial conditions. In particular, initial datum has been taken to be a Poisson random variable with mean value $\mu$. Additionally, the case of an initial distribution to be Gaussian and with mean value $\mu$ and variance 0.5$\mu$ and 1.5$\mu$ has been considered. The mean value $\mu$ has been set to $\lambda$z for sub--lethal lesions and $\kappa$z for lethal lesions. It can be seen how the initial value can sharpen or broaden lethal and sub--lethal lesion distributions, with a straightforward consequence on the resulting survival probability, see figure \ref{FIG:Survival}.

Survival probability is one of the most used and relevant radiobiological endpoints. Figure \ref{FIG:Survival} highlights how a different initial condition affects the resulting survival curve. 
In particular, it is important to notice that the probability of survival rises or falls in the high dose region. One of the major flaws in classical models, with particular reference to the linear--quadratic model, is the fact that it significantly underestimates the probability of survival for high doses.

\section{Conclusion}

The present work represents a first step into an advanced and systematic investigation of the stochastic nature of energy deposition by particle beams, with particular focus on how it affects DNA damage. Starting from  basic probabilistic assumptions, a \textit{master equation} for the probability distribution of the number of lethal and sub--lethal lesions induced by radiation of a cell nucleus has been derived. The new model, called (\textit{Generalized Stochastic Microdosimetric Model} (GSM$^2$), provides a simple and yet fundamental generalization of all existing models for DNA-damage prediction, being able to truly describe the stochastic nature of energy deposition. This advance results in a more general description of DNA-damage formation and time-evolution in a cell nucleus for different irradiation scenarios, from which radiobiological outcomes can be assessed.

Most of the existing models assume a Poissonian distribution of lethal damage, ignoring the true space-time stochastic nature of energy deposition. 
In order to overcome the limits of this assumption, \textit{ad hoc} corrections have been introduced, called non-Poissonian corrections in the literature, but an extensive survey on the complete stochasticity of biophysical processes to the best of our knowledge has never been carried out.

This work aims at highlighting how the stochastic nature of energy deposition can lead to different cell survival estimations and how non--Poissonian effects emerge naturally in the general setting developed. Remarkably, in particular, GSM$^2$ does not require any \textit{ad hoc} corrections for taking into account overkill effects, as it is required by all existing models.

A further investigation, dedicated to a separate work, will focus on verification and optimization of the prediction of the survival curves for different systems, i.e. radiation type, irradiation conditions and cell line. In addition, given the general nature of the proposed model, closed form solutions for lesion distribution and survival curve are typically difficult to obtain. However, it is fair to say that, due to the several process involved, approximation methods provide powerful tools to estimates several quantity of interests. Among the most important approximation methods, we mention system size expansions \cite{Kur,VK,Gar,Mel}, and the related small--noise asymptotic expansions \cite{Gar}. Both approaches will be investigated in future research to provide accurate estimates of several biological endpoints, such as cell survival.

Further investigation will also be devoted to develop a more efficient numerical implementation of the driving \textit{master equation}.

\section{Acknowledgments}
This work was partially supported by the INFN CSNV projects MoVe-IT and NEPTune.

\cleardoublepage
\bibliographystyle{apalike}

\bibliography{bib}

\end{document}